\documentclass[iop]{emulateapj}

\def \mum{$\mu$m}

\shorttitle{NGC 1333 mid-IR Proper Motions}
\shortauthors{Raga, Noriega-Crespo, Carey \& Arce}

\begin{document}
\def \mum{$\mu$m}


\title{Proper Motions of Young Stellar Outflows in the Mid-Infrared with
Spitzer (IRAC). I. The NGC 1333 region}


\author{A. C. Raga\altaffilmark{1}}

\affil{$^1$
Instituto de Ciencias Nucleares, UNAM, Ap. 70-543, 04510 D.F., Mexico}
\email{raga@nucleares.unam.mx}

\author{A. Noriega-Crespo\altaffilmark{2}}

\affil{$^2$ Infrared Processing and Analysis Center, California Institute of
Technology, Pasadena, CA 91125, USA}
\email{alberto@ipac.caltech.edu}

\author{S. J. Carey\altaffilmark{3}}

\affil{$^1$Spitzer Science Center, California Institute of
Technology, Pasadena, CA 91125, USA}
\email{carey@ipac.caltech.edu}

\author{H. G. Arce\altaffilmark{4}}

\affil{$^4$
Department of Astronomy, Yale University, New Haven, CT 06520, USA}
\email{hector.arce@yale.edu}

\begin{abstract}
We use two 4.5\mum~Spitzer (IRAC) maps of the NGC~1333 region taken over
a $\sim 7$ yr interval to determine proper motions of its associated outflows. 
This is a first, successful attempt at obtaining proper motions of stellars 
outflow from Spitzer observations.
For the outflow formed by the Herbig-Haro objects HH7, 8 and 10, we find
proper motions of $\sim 9$-13 km s$^{-1}$, which are consistent with
previously determined optical proper motions of these objects. We
We determine proper motions for
a total of 8 outflows, ranging from $\sim 10$ to 100 km s$^{-1}$.
The derived proper motions show that out of these 8 outflows,
3 have tangential velocities $\leq 20$ km s$^{-1}$. This result shows
that a large fraction of the observed outflows have low intrinsic
velocities, and that the low proper motions are not merely a projection
effect.
\end{abstract}



\section{Introduction}

Starting with the work of Herbig \& Jones (1981), who studied the
HH 1/2 outflow, the study of proper
motions of Herbig-Haro (HH) objects has become one of the fundamental
tools in exploring the properties of outflows from young stars. 
Proper motions of young stellar outflows were first measured
at radio wavelengths by Rodr\'\i guez et al. (1989), who
studied the Serpens ``triple source''.
Later, Noriega-Crespo et al. (1997) first obtained proper motions 
of outflows at IR wavelengths of HH1, from H$_2$ 2.12$\mu$m images.

In contrast to what has happened at optical and radio wavelengths,
the study of IR proper motions of young stellar outflows has not
flourished. This is probably due to the fact that older IR images typically
cover only relatively small fields, so that it is many times
impossible to obtain a good centering/scaling between them and
newer (larger field) IR images. Previous IR proper motion studies
include Noriega-Crespo et al. (1997: HH1), Chrysostomou et al.
(2000: HH7-11, 25-26 and 33/40) and Raines et al. (2000: GGD37).

In this paper we make a first attempt to measure proper motions
from Spitzer-IRAC IR images of young stellar outflows. This is
possible because even though the IRAC pixels are relatively
large ($1''.2$), some outflows have now been observed over time
intervals of $\sim 10$~yr. For example, for an outflow with
plane-of-the-sky velocities of $\sim 100$~km~s$^{-1}$ at
a distance of $\sim 500$~pc, offsets of $\sim 0.35$ pixels
are expected over $\sim 10$~yr. As shown initially by Raga, Barnes
\& Mateo (1990), such offsets can be measured with reasonable
accuracy in CCD images of stellar outflows.

There are three fundamental reasons that allow us to use IRAC
to measure proper motions in the mid-IR, despite its relative low
angular resolution (FWHM$\sim$2\arcsec) in comparison
with optical or near-IR ground based observations. First of all
is the outstanding stability of the pointing control system of the
Spitzer Space Telescope over the past nine years.
Secondly, the unchanged performance of its optical system and
its detectors. Thirdly, the ability of IRAC to map large regions of
the sky, thus providing plenty of point sources that can be used to
create and fix a reliable reference frame to compare
observations from different epochs.

We study two 4.5\mum~IRAC frames of the NGC~1333 region,
obtained in February/September 2004 and October/November 2011.
These observations are described in detail in section 2.

These images allow us to obtain
the proper motions of some of the HH7-11 knots (HH7, 8
and 10), which can be compared with previous optical
(Herbig \& Jones 1983; Noriega-Crespo \& Garnavich 2001)
and IR (Chrysostomou et al. 2000) proper motion measurements
of this outflow. This is done in section 3. Also, these images
allow us to obtain the proper motions of other outflows in
NGC~1333, as described in section 4. Finally, our results are
summarized in section 5.

\section{Observations}

NGC~1333 was one of the first star forming regions observed with
the Spitzer Space Telescope (Werner et al. 2004) with the
Infrared Array Camera (IRAC) (Fazio et al. 2004)
during the Cryo-mission as part of the IRAC GTO program of
its principal investigator (PI) (PID 6, G. Fazio). This first epoch
was complemented approximately six months later by 
the  Cores-to-Disks Spitzer Legacy program as part of their
mapping of the Perseus molecular cloud  (PID 178, PI N. Evans).

Multiple epoch observations of star forming regions relatively near
the ecliptical plane are essential to weed out asteroids when trying
to identify real proto-stellar objects (see e.g. Evans et al. 2009).
The more recent data were obtained in October-November 2011, 
and are being currently analyzed to study the
time variability of young stellar objects (YSOs) in the region,
as part of the Warm-mission Spitzer Exploration program YSOVAR
(PID 61026, PI J. Stauffer). YSOVAR has observed the core of NGC~1333,
a $\sim$ 16\arcmin$\times$11\arcmin~region
centered on 03h29m10.175s +31d16m16.0s, over a period of nearly a month
using 73 different visits. Each visit lasted  approximately 15 minutes
with a mean coverage of $\sim 4 - 5$ in the final maps,
although some pixels in the overlap regions
can double these values.
Our final 2010 image has a mean coverage of $\sim 300$, i.e., on average
each pixel has seen the object some 300 times, and since the observing time
per pixel is 12 sec, this corresponds to a $\sim$ one hour total exposure
time.

This central map has determined the size of the region that we have
used for our analysis of the proper motions between the two (2004 \& 2011)
available epochs.  The final map is actually a bit smaller to avoid
the scattered light present around the very bright young star LZK 12
in the very deep YSOVAR image. The Cryo-mission observations were
carried using all IRAC channels (3.6, 4.5, 5.8 \& 8.0\mum), while in
the Warm-mission observations we can only use channels 1 and 2
(3.6 and 4.5\mum). For the three mentioned programs,
the data was collected using the High-Dynamic-Range (HDR) mode with
a 12sec integration time for the 'long' frames and 0.6sec 
for the 'short' ones. For the purpose of our analysis we have
concentrated on the data at 4.5\mum, which for young stellar outflows has
proven to be a very good tracer of the
molecular Hydrogen emission of some of the bright pure rotational transitions
[(0-0 S(9) at 4.6947, 0-0 S(10) at 4.4096 and 0-0 S(11) at 4.1810\mum],
and therefore, of their shock excited emission (Noriega-Crespo et al. 2004a,b;
Looney et al. 2007; Tobin et al. 2007; Ybarra \& Lada 2009; Maret et al.
2009; Raga et al. 2011; Noriega-Crespo \& Raga 2012).

For the IRAC observations we have used the latest versions
of the data, S18.18.0 (Cryo) and  S19.0.0 (Warm),
and a scale of 0.3\arcsec/pix for the final maps, i.~e.
a seventh of the IRAC FWHM$\sim$ 2\arcsec~angular resolution.
The maps of the different epochs are created with 
the same fiducial frame, using MOPEX (Makovoz, Khan \& Masci 2006), 
which brings them very close to the same frame of reference.
A dozen non-saturated stars common between the two epochs are
then used for the final alignment of the images, with a typical
centroid accuracy of less than a tenth of a pixel. The two resulting
maps (2004 and 2011) are shown in Figure 1.

\begin{figure}
\centering
\includegraphics[width=1.1\columnwidth,angle=0]{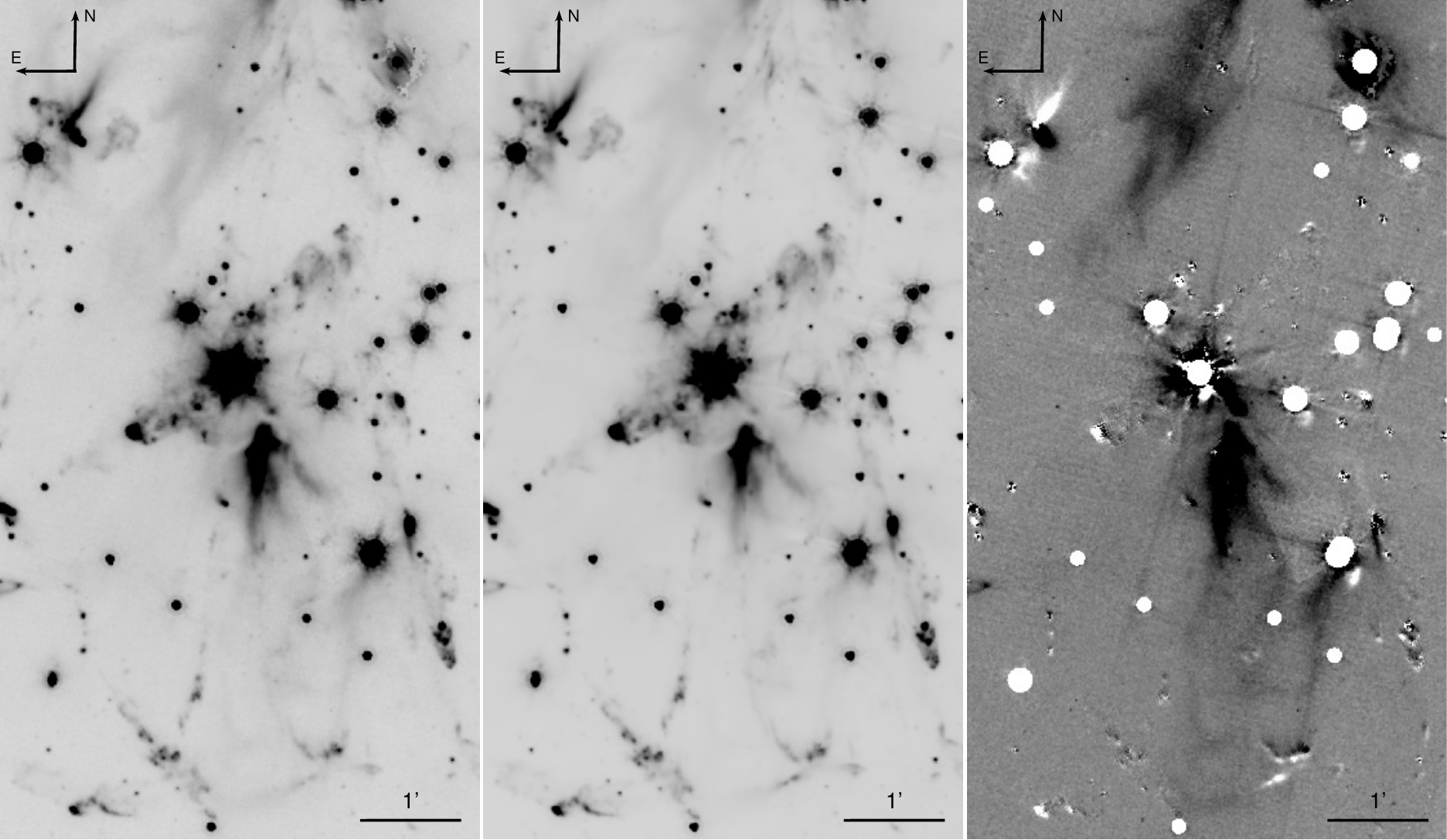}
\caption{Top frame: $2004-2011$ difference image.
Center: 2011 image. Bottom: 2004 image. In the difference image,
the knots along outflows with proper motions show up
as ``positive/negative'' double ridge structures, and the white
disks are masked stars.
\label{fig1}}
\end{figure}

Figure 1 shows the two final images (2004 and 2011), 
and the difference between these two images. In the difference map, 
one can clearly see that some of the compact emission features show 
clear proper motions. For example, a bow-shaped region in
the SW region of the frame shows a ``positive/negative''
ridge combination which clearly indicates a motion of this
feature. Such motions are analyzed in detail below.

Although there are no major changes in the optics of the telescope
or the performance of the detectors between the IRAC Cryo and Warm
missions, a comparison between the
2004 and 2011 images shows stronger optical ghosts around 
the bright sources in the Warm data. This effect is probably
enhanced because of the very deep coverage of the YSOVAR data (with
$\sim$ 300 exposures of 12 sec for each pixel, see above).
Figure 2 shows a 3.5\arcmin$\times$3.5\arcmin region of NGC~1333
with some bright sources, where one can see the typical 'heart shape'
IRAC Point Spread Function, with the nearby
``triangular ghosts'' marked with the white boxes (12\arcsec$\times$12\arcsec).
These are an annoyance more than a problem in measuring the proper
motions, since they appear as real time variations between epochs.
For this reason we have masked some of them in both final (2004 and 2011)
images. The observations are summarized in Table 1.

\begin{figure}
\centering
\includegraphics[width=0.8\columnwidth]{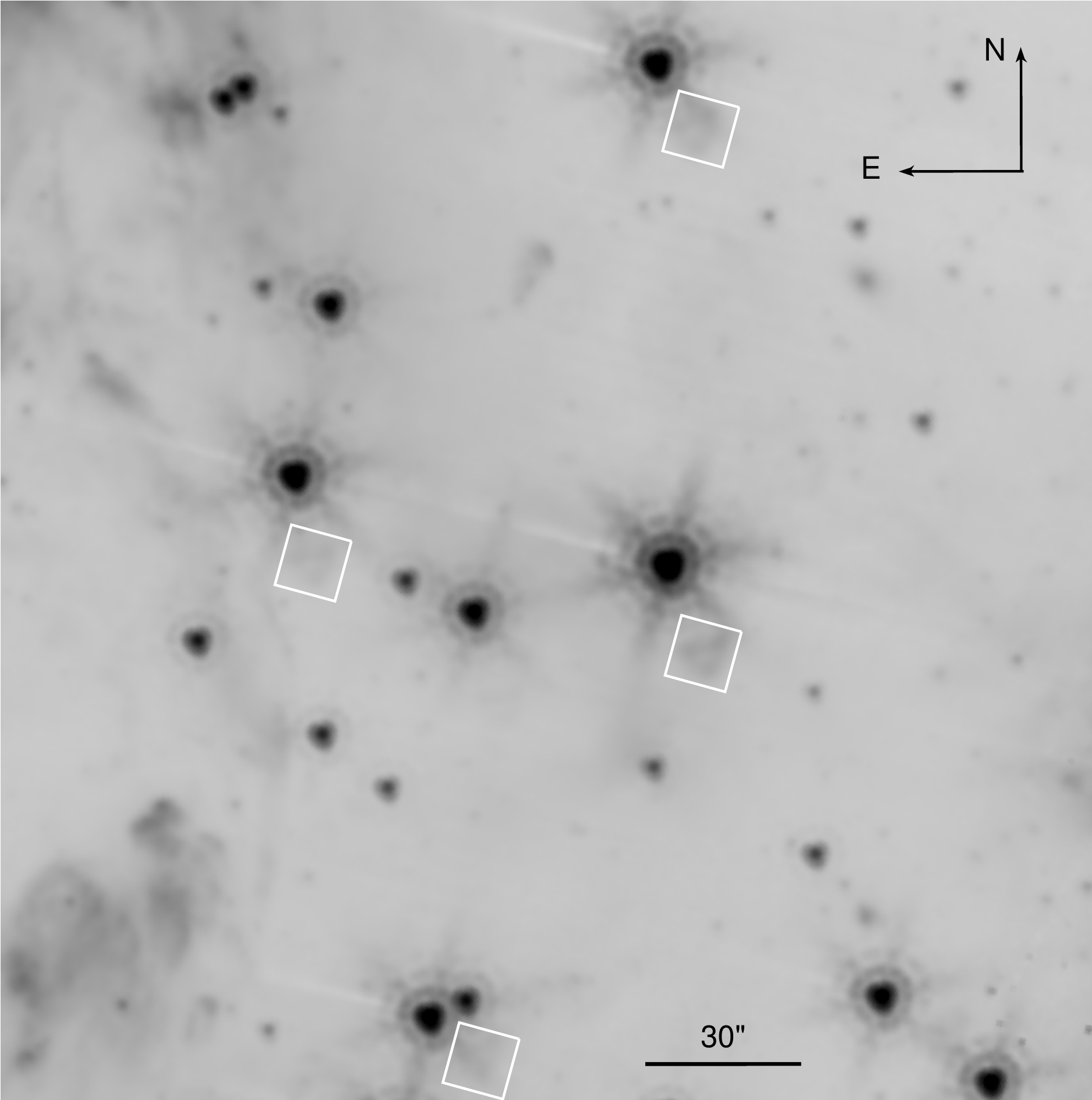}
\caption{Region of the ``Warm'' (2011) analyzed frame. The white boxes
indicate the ``ghost'' structure associated with the PSF in this frame
(see section 2).
\label{fig2}}
\end{figure}

\section{The HH7-11 outflow}

In Figure 3, we show the 2004 frame of the HH7-11 outflow.
HH 9 is not seen in the 4.5\mum~IRAC maps, and HH11
is lost within the PSF of SVS13.
The image shows a knot $\approx 20''$ to the S of HH10, which
we have labeled ``knot A'' (see Figure 3).

Figure 3 also shows proper motion vectors calculated with the ``proper
motion mapping'' technique described by Raga et al. (2012). This
(rather simplistic) technique is described in Appendix A.
In particular, the HH7-11 proper motions shown in Figure 3 were
computed with cross correlations in boxes of size $L=30$ pixels
(with each pixel corresponding to $0''.3$, see section 2) which
have peak fluxes of at least $f_{min}=2.0$ mJy/sterad (see
Appendix A). The proper motion velocities were calculated
assuming a distance of 220~pc to NGC~1333 ({\~ C}ernis 1990;
Hirota et al. 2008).

\begin{figure}
\centering
\includegraphics[width=0.8\columnwidth]{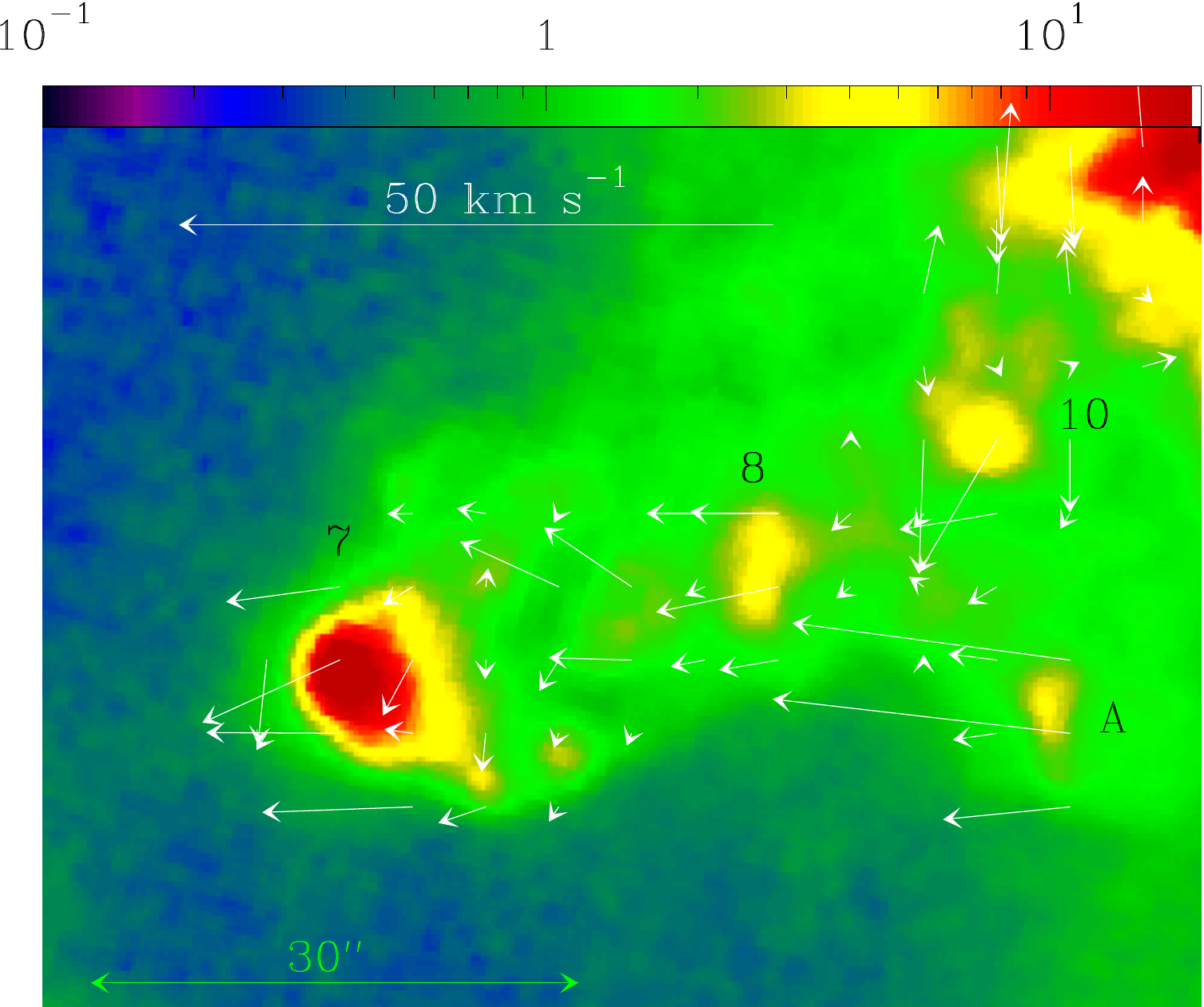}
\caption{The map shows HH7-11 outflow lobe in the 2004 map.
The white arrows indicate the proper motion velocities
(for a distance of 220 pc to the system)
computed within the cross correlation boxes with high
enough intensities. We detect motions
for HH 7, 8 and 10 (HH 11 being confused with the
stellar PFS off the NW corner of the frame).
N is up and E to the left. The image is displayed with
a logarithmic scale given (in mJy/sterad) by the top bar.
\label{fig3}}
\end{figure}

The HH7, 8 and 10 knots have velocities $\approx (10\pm 4)$~km~s$^{-1}$
(HH7: 13 km s$^{-1}$; HH8: 10 km s$^{-1}$; HH10: 9 km s$^{-1}$),
with HH7 and HH8 moving approximately parallel to the outflow
axis, and HH10 showing a possibly significant deviation towards
the S. These proper motions were obtained by averaging the
proper motions from the two boxes closest to the peak emisison
of the knots. Also the region between HH7 and 8 shows filamentary
structures which apparently share the kinematics of these
two HH knots (see Figure 3).

Knot A has a significantly higher proper motion velocity
of $(25\pm 4)$ km s$^{-1}$, pointing approximately E. This
knot could belong to one of the other outflows present
in this complex region (see section 4 and Davis et al. 2011).

Finally, relatively large proper motions are seen within
$\sim 20''$ from the SVS13 outflow source (outside the upper,
right hand side corner of Figure 3), which are probably related
with the spikes of the PSF, and are unlikely to represent
real motions in the flow.

The proper motions that we have obtained for HH 7, 8 and
10 are roughly consistent with the velocities of
$\sim (20 \pm 10)$~km~s$^{-1}$ determined for these knots with
the proper motion determinations of Noriega-Crespo \& Garnavich (2001),
scaled to a distance of 220~pc to HH7-11. The proper motions
of Herbig \& Jones (1981) are also consistent with these relatively
low velocities. The velocities of $\sim 400$~km~s$^{-1}$
determined by Chrysostomou et al. (2000) are inconsistent with the
three other independent (optical and IR) proper motion determinations.

\section{Other outflows}

In Figure 4, we show a $450\times 550''$ field (covered in the
two epochs) approximately centered on the HH 7-11 outflow. On the
2004 frame, we show the axes of the molecular outflows (which
we have numbered 1 through 9) described
by Davis et al. (2008). As already seen in the diffenerce
image between the two epochs (see Figure 1 and section 2),
there are several regions of extended emission which show
detectable proper motions.

The proper motions shown in Figure 4
correspond to cross correlations within $L=60$ ($18''$) pixel boxes
of the two epochs, with fluxes of at
least $f_{min}=0.9$ mJy/sterad (see Appendix A).

We have chosen 6 regions (labeled
A through F in Figure 4) which show the more interesting
proper motion detections. These regions are described in the following
subsections.

\begin{figure}
\centering
\includegraphics[width=0.8\columnwidth]{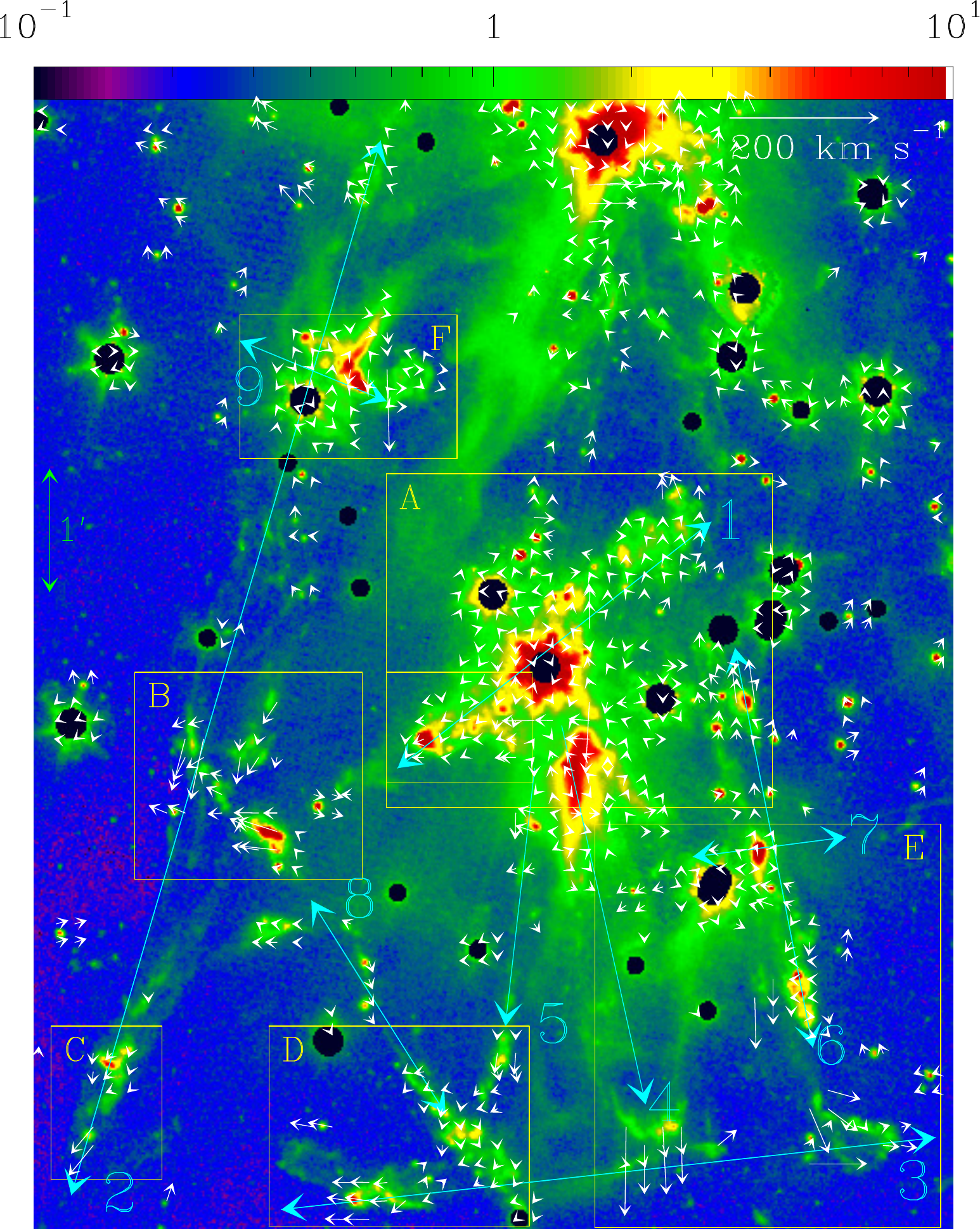}
\caption{2004 frame (displayed with the logarithmic
scale given in mJy/sterad by the top bar) of the studied
field. The dark, circular regions correspond to masked stars.
The blue arrows indicate the axes of the outflows
described by Davis et al. (2008), which
we have numbered 1 through 9. The yellow boxes (labeled
A through F) indicate interesting subfields, which are
shown in an expanded scale in Figures 5-10. The white
arrows indicate the proper motion velocities (for
a distance of 220 pc) obtained through our ``cross
correlation mapping'' technique. N is up and E to the left.
\label{fig4}}
\end{figure}

\subsection{Region A} 

This region (see Figure 5)
includes the HH 7-11 outflow (the smaller box
within region A in Figure 4, corresponding to the blue-shifted, 7-11 lobe,
is shown in Figure 3). It is clear that while organized proper motions (aligned
with the outflow axis) are seen along the HH 7, 8, 10 knots, such
motions are not seen in the NW outflow lobe.

To the S of the SVS 13
source, relatively large proper motions with Easterly direction
are seen, but it is not clear that these motions are associated
with well defined knots (except for ``knot A'' of Figure 3, see
section 3). These motions might be associated with a previously
undetected outflow.

In the center of the frame, at $\sim 30''$ from
the Southern edge of the frame, we see
a bow-like region with Northward proper motions of
$\sim 10$~km~s$^{-1}$. These marginally detected motions
could be associated with the Northern lobe of outflow 4 (see Figure 4).

\begin{figure}
\centering
\includegraphics[width=0.8\columnwidth]{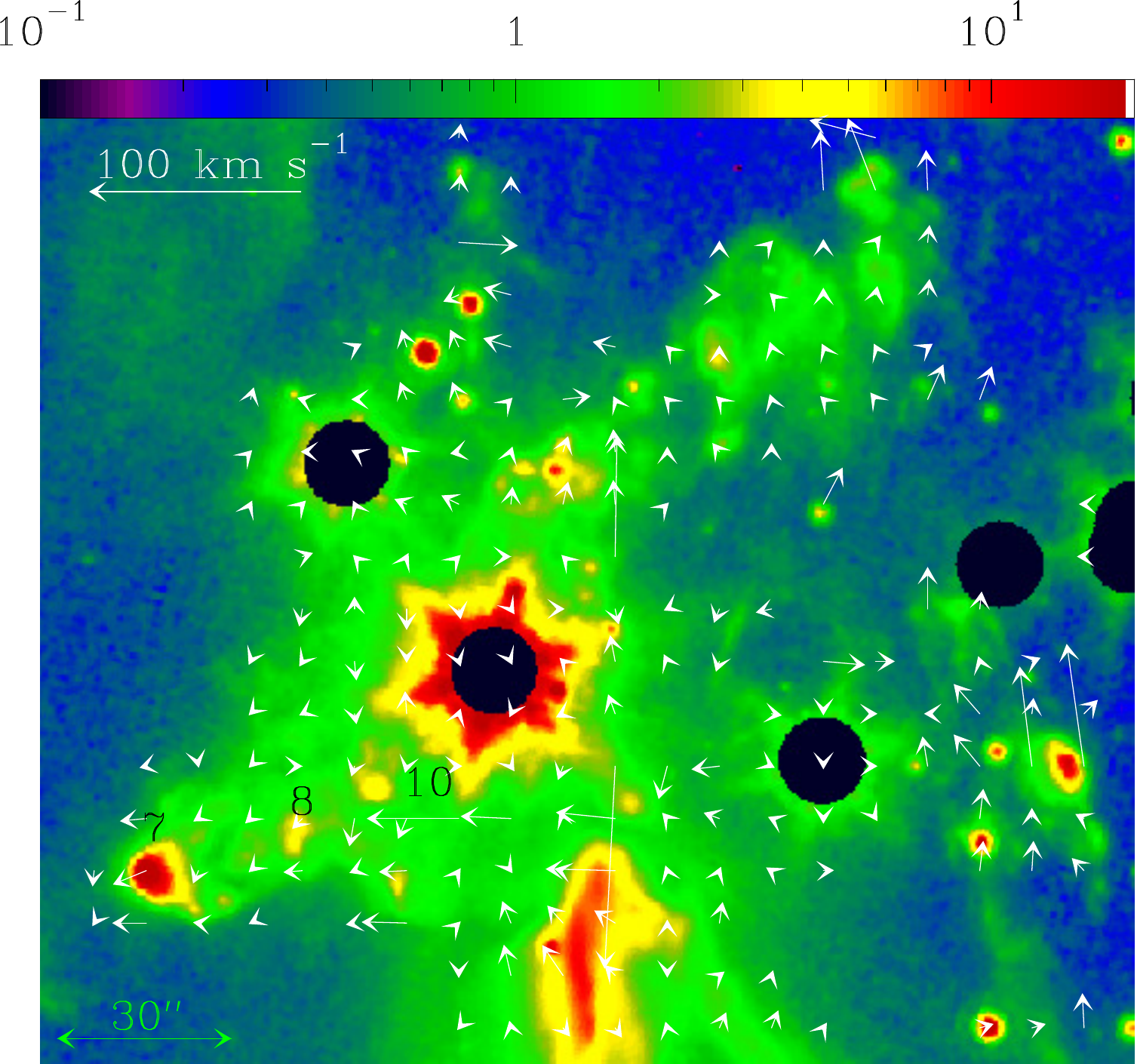}
\caption{Subfield A of Figure 4. The 2004 map is shown
together with the proper motion velocities.
N is up and E to the left.
\label{fig5}}
\end{figure}

Close to the SW corner of the frame, we see a region
with a broken-up, bow-like structure with
proper motion velocities of $\sim 20$-70 km s$^{-1}$
to the N or NE. This motion
appears to be shared by some of the knots seen in the NW
region of Figure 5. All of these Northward directed knots appear
to belong to outflow 6 (see Figure 4).

\subsection{Region B}

This region (see Figures 4 and 6) shows two
systems of knots with different proper motions. One set of
knots (in the Northern part of Figure 6) has proper motion
velocities of 20-100 km s$^{-1}$ directed approximately
to the SSE. This set of knots appears to be associated
with outflow 2 (see Figure 4).

A second set of knots (in the Southern half of
Figure 6) shows proper motions of similar magnitudes
(10-60 km s$^{-1}$), but directed Eastwards. These
knots could be associated with outflow 8 (see Figure 4),
though their motion would imply a substantial
deflection from the outflow axis. Alternatively,
these knots could belong to a previously undetected outflow.

\begin{figure}
\centering
\includegraphics[width=0.8\columnwidth]{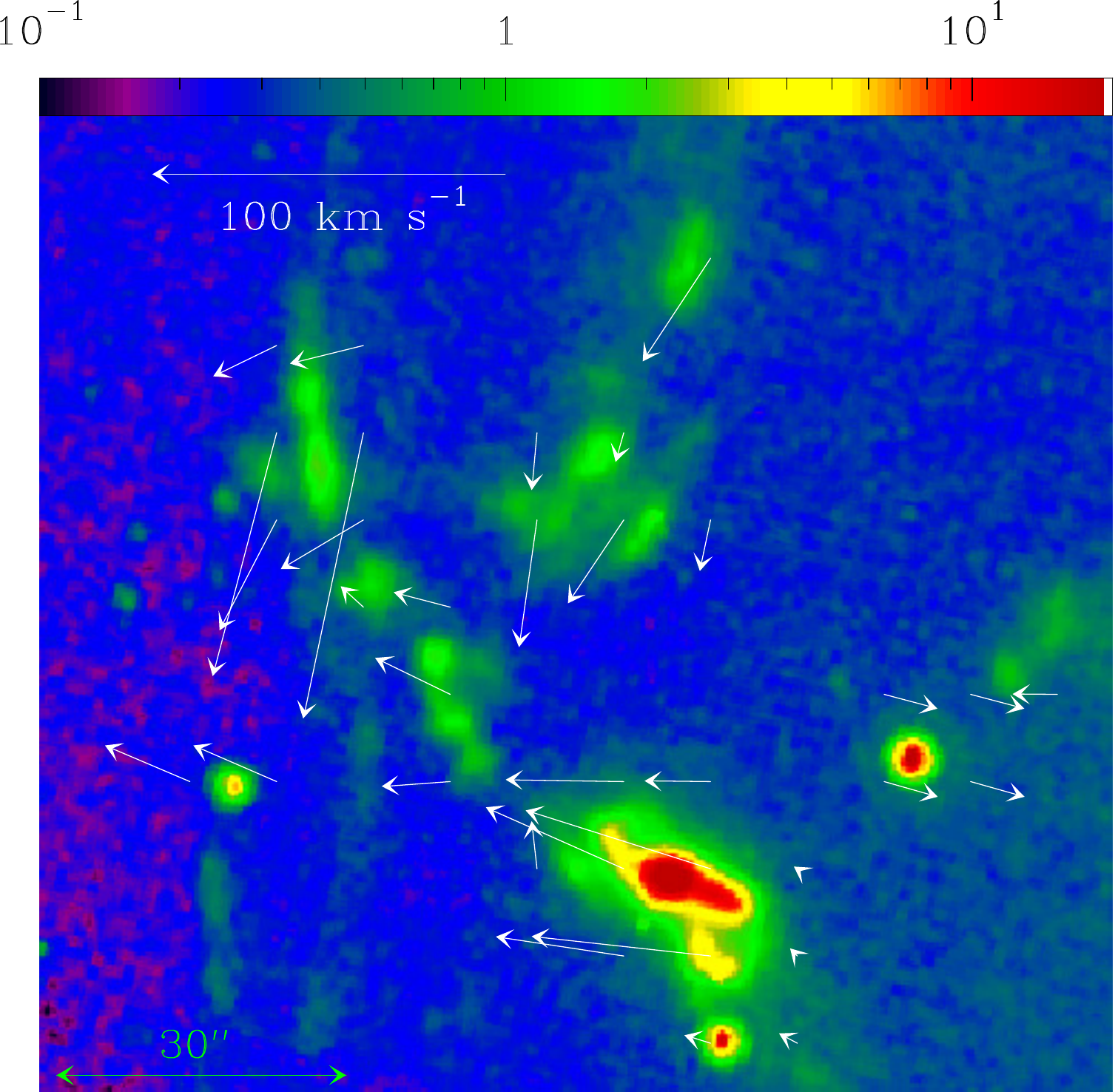}
\caption{Subfield B of Figure 4. The 2004 map is shown
together with the proper motion velocities.
N is up and E to the left.
\label{fig6}}
\end{figure}

\subsection{Region C}

This region (see Figures 4 and 7) shows
knots with proper motions of 10-60 km s$^{-1}$ directed towards
the SE. These knots are therefore likely to be a prolongation
of the S lobe of outflow 2 (see Figure 4).

\subsection{Region D}

The Southern part of this region (Figure 8) shows
a bow-like structure with Eastwards proper motions
of 20-40 km s$^{-1}$, apparently associated with
the E lobe of outflow 3 (see Figure 4). In the NW
corner of the frame, we find a set of knots with
Southward proper motions of 20-40 km s$^{-1}$, which
are likely to be associated with outflow 5 (see Figure 4).

The region which is morphologically aligned with
outflow 8 (extending from the N center to the SW corner
of Figure 8) does not show proper motions aligned
with the outflow axis.

\subsection{Region E}

In the SW corner
of this region (Figure 9) we see a set of knots with
Western proper motions of 10-80 km s$^{-1}$ which
are morphologically and kinematically associated with
the W lobe of outflow 3 (see Figure 4).

In the SE
corner, we see a bow shaped structure with Southern proper motions
of up to 100 km s$^{-1}$, which is likely to be
associated with outflow 4. In the
center of this region we see knots with
Southern proper motions of $\sim 60$ km s$^{-1}$
which might be associated with outflow 6 (see
Figure 4).

Finally, in the NE corner of this field (see Figure 9) we see
aligned proper motion vectors towards the E, corresponding
to velocities $\sim 15$~km~s$^{-1}$. These motions
could be associated with the Eastern lobe of outflow 7
(see Figure 4).

\begin{figure}
\centering
\includegraphics[width=0.8\columnwidth]{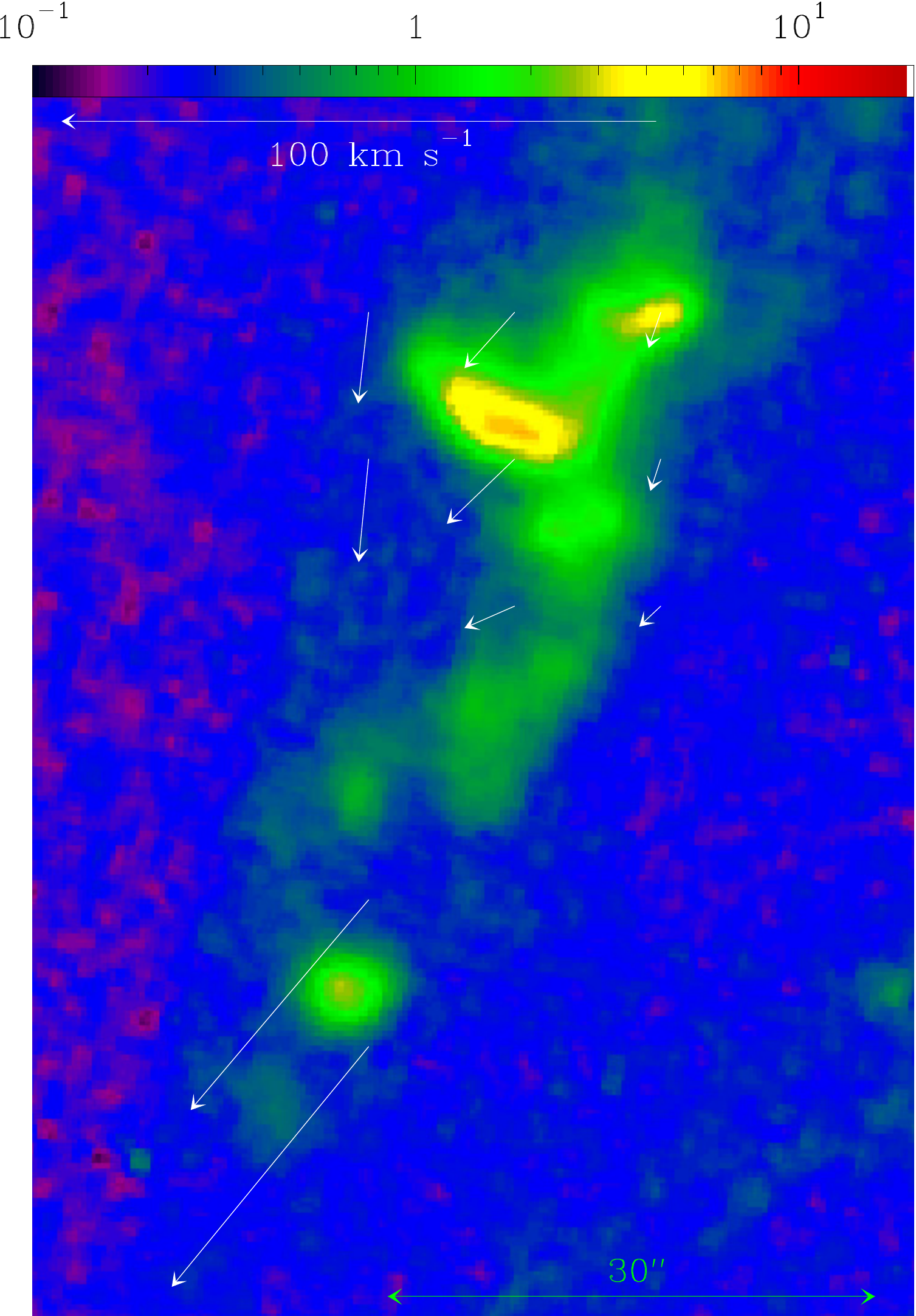}
\caption{Subfield C of Figure 4. The 2004 map is shown
together with the proper motion velocities.
N is up and E to the left.
\label{fig7}}
\end{figure}

\begin{figure}
\centering
\includegraphics[width=0.8\columnwidth]{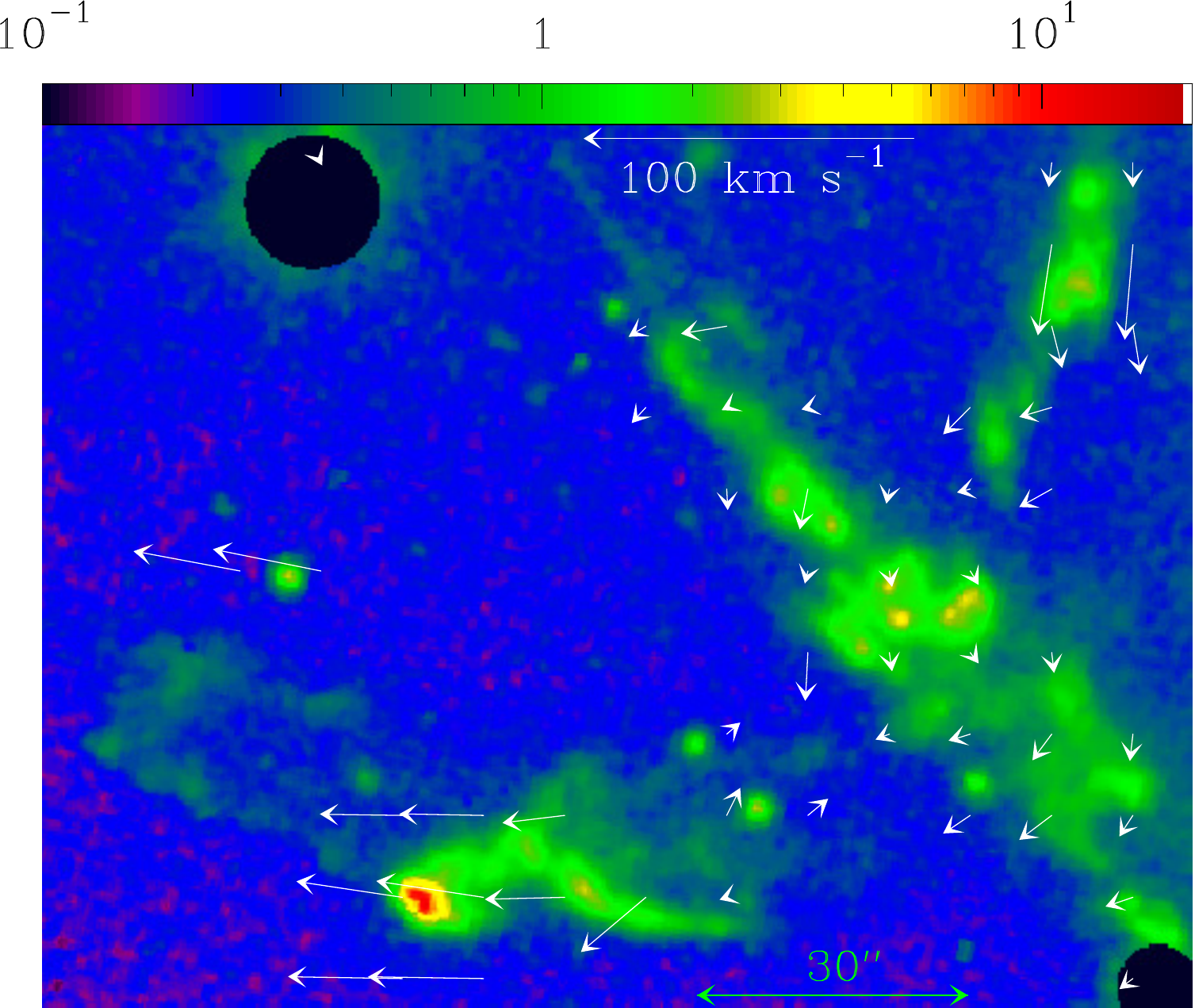}
\caption{Subfield D of Figure 4. The 2004 map is shown
together with the proper motion velocities.
N is up and E to the left.
\label{fig8}}
\end{figure}

\begin{figure}
\centering
\includegraphics[width=0.8\columnwidth]{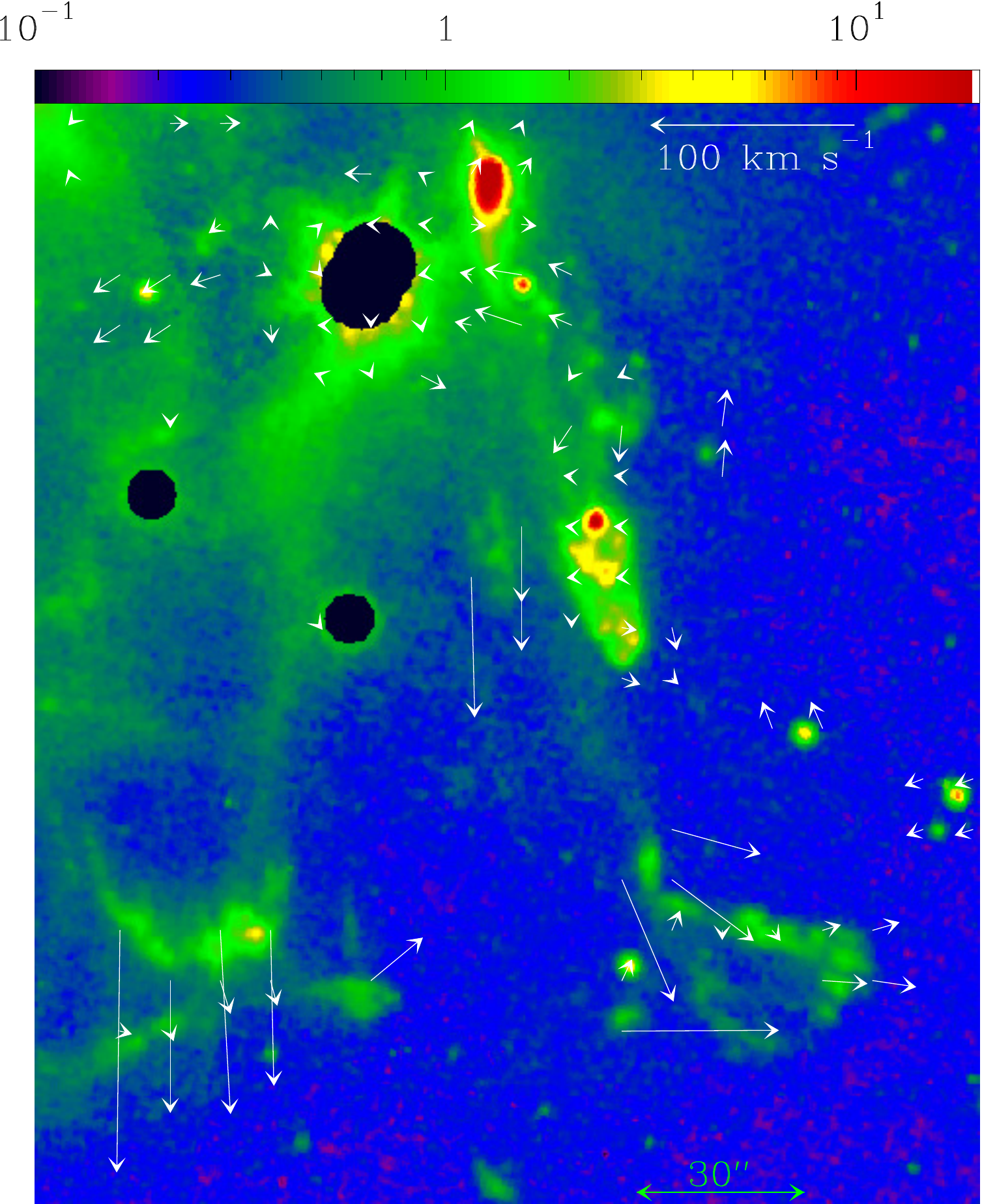}
\caption{Subfield E of Figure 4. The 2004 map is shown
together with the proper motion velocities.
N is up and E to the left.
\label{fig9}}
\end{figure}

\subsection{Region F}

This region (Figure 10)
shows some high Southwards proper motions
of $\sim 50$~km~s$^{-1}$ (actually, one of the
Southwards proper motion velocities
is $>100$ km s$^{-1}$, but for a cross-correlation
box without any clear feature), which do not have
a clear identification with the outflows of Davis
et al. (2008).

Also, this region might be showing evidence of
a bipolar structure (with oppositely directed
proper motion velocities of $\sim 10$ km s$^{-1}$, indicated
with two open circles in Figure 10), which
might be associated with outflow 9 (see Figure 4).

\subsection{Proper motions for the outflows of Davis et al. (2008)}

As described above, we detect proper motions for knots which are
likely to be associated with several of the outflows of
Davis et al. (2008). These outflows are numbered 1 through 9 in Figure 4.

For each of these outflows, the following results are obtained:
\begin{itemize}
\item {\it outflow 1~-} we measure proper motions of up to
20 km s$^{-1}$ along the SE lobe of this outflow. These proper motions
are consistent with the optical proper motions measured for
HH 7-11 by Noriega-Crespo \& Garnavich (2001). We do not
see organized motions in the region that would correspond
to the NW lobe of this outflow,
\item {\it outflow 2~-} we detect two sets of knots (within
regions B and C of Figure 4, respectively) with
proper motions of up to $\sim 100$ km s$^{-1}$ aligned with
the SE lobe of this outflow,
\item {\it outflow 3~-} we detect proper motions along both the
E (region D) and W (region E) lobes of this outflow, with
velocities of up to $\sim 80$ km s$^{-1}$,
\item {\it outflow 4~-} we detect a Southern bow shock with
proper motions of $\sim 100$ km s$^{-1}$ (region E) and
possibly a Northern bow shock with much lower ($\sim 10$ km s$^{-1}$)
velocities (region A), which might be associated with this outflow,
\item {\it outflow 5~-} we detect knots with velocities of
$\sim 30$ km s$^{-1}$ which might be associated with the Southern
lobe of this flow (region D),
\item {\it outflow 6~-} we detect knots with proper motions of
$\sim 50$ km s$^{-1}$ which are likely to be associated with
the Northern (region A) and Southern (region E) lobes of this outflow,
\item {\it outflow 7~-} the E lobe of this outflow coincides
with IR knots with velocities of $\sim 15$ km s$^{-1}$,
\item {\it outflow 8~-} we do not obtain proper motion velocities
aligned with this outflow,
\item {\it outflow 9~-} we marginally detect
proper motion velocities of $\sim 10$ km s$^{-1}$  which
might be associated with this outflow.
\end{itemize}
Therefore, we detect knots with measurable proper motions which
are likely to be associated with all but one of the outflows
of Davis et al. (2008).

\begin{figure}
\centering
\includegraphics[width=0.8\columnwidth]{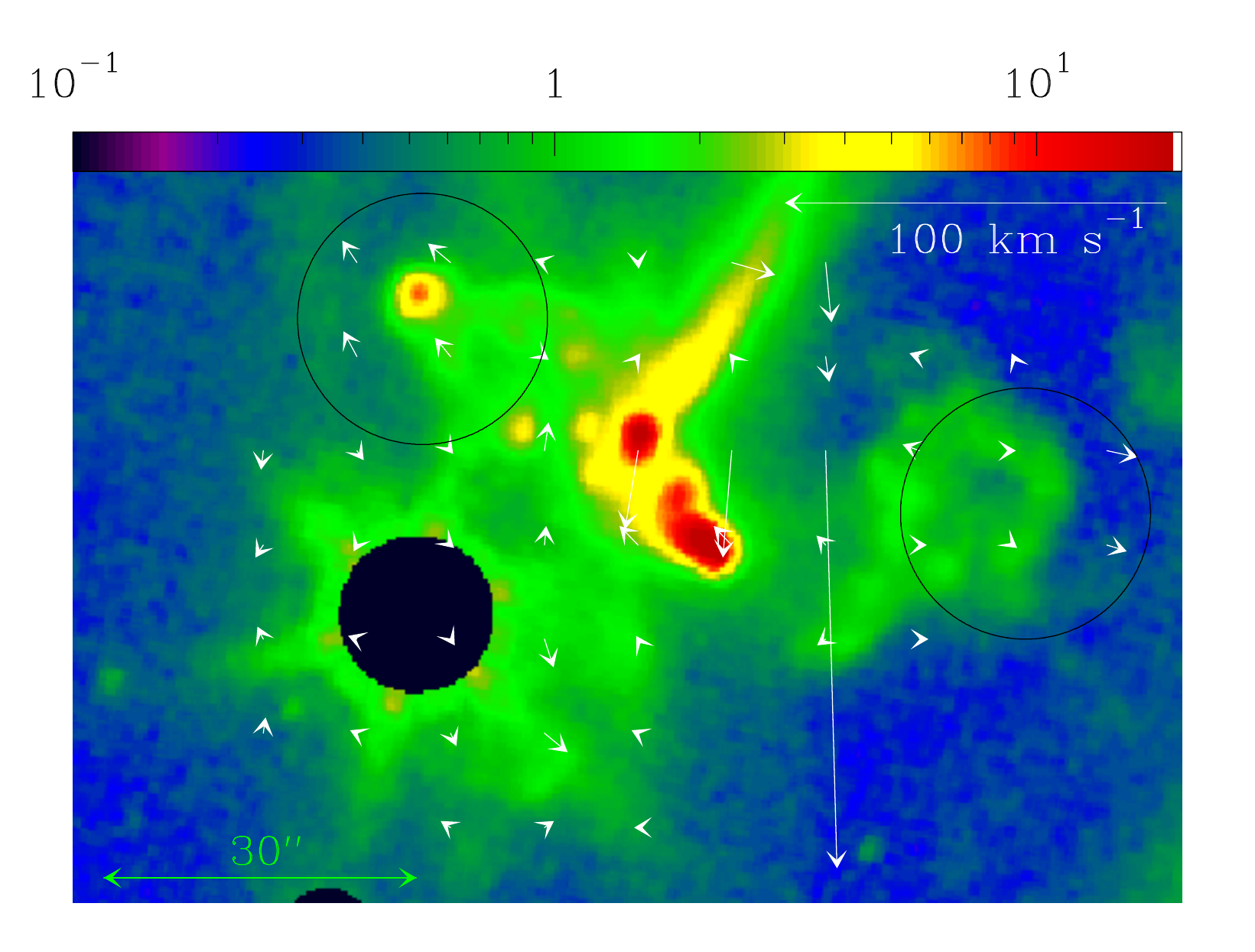}
\caption{Subfield F of Figure 4. The 2004 map is shown
together with the proper motion velocities. The two open
circles indicate the regions that could be associated
with the two lobes of outflow 9 (see section 4.6).
N is up and E to the left.
\label{fig10}}
\end{figure}

\subsection{The proper motion velocity distribution function}

If we are detecting most of the outflows in NGC 1333, the measured
proper motion velocities $v_T$ allow us to obtain the
distribution function associated with the outflows from a group
of stars currently being formed. If we consider the maximum value of
$v_T$ for the 8 outflows of Davis et al. (2008) with measured
proper motions (see subsection 4.7), we find that we have
3 with maximum $v_T$ in the $0\to 20$ km s$^{-1}$ range,
1 outflow in each of the $20\to 40$, $40\to 60$ and $60\to 80$ km s$^{-1}$
ranges, and 2 outflows with $v_T$ in the $80\to 100$ km s$^{-1}$ range.
The resulting distribution function (normalized to a total of 5 outflows,
see below) is shown in Figure 11.

The distribution function shows an apparent peak at low proper motion
velocities ($v_T$), and possibly also a marginal peak in the higher
velocity bin. One can show that this distribution is not consistent
with a set of outflows with identical intrinsic velocities $v_j$
and with arbitrary orientation angles with respect to the plane
of the sky.

It is straightforward to show that if we have a set of outflows
with outflow velocity $v_j$ and with arbitrary orientations, the expected
distribution function for the proper motion velocities $v_T$ is:
\begin{equation}
f(v_T)=\frac{1}{v_j}\left[\left(\frac{v_j}{v_T}\right)^2-1\right]^{-1/2}
\label{fv}
\end{equation}
This distribution function is plotted in Figure 11, considering
$v_j=100$ km s$^{-1}$. In this Figure, the
distribution function obtained from the 8 observed outflows has been
normalized to 5 outflows, i.e., not considering the 3 outflows in the
$0\to 20$ km s$^{-1}$ bin. A comparison between the observed and
theoretical distributions shows that a single population of
$v_j=100$ km s$^{-1}$ outflows (at random orientations) does not produce
a proper motion velocity distribution consistent with the data.

\begin{figure}
\centerline{
\includegraphics[width=200pt,height=200pt,angle=0]{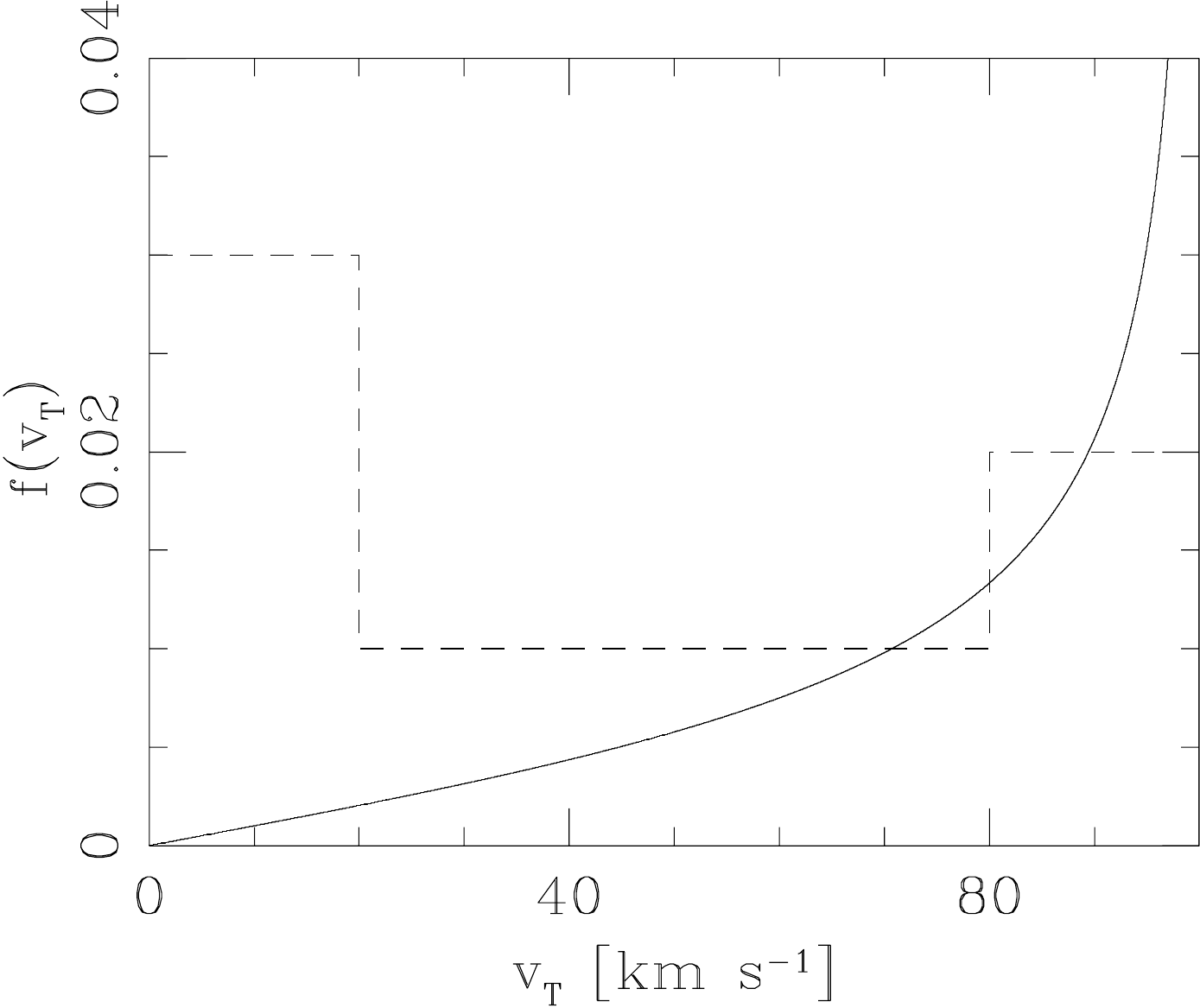}}
\caption{\label{fig11}The dashed histogram shows the
distribution function obtained by binning the
peak proper motion velocities of the 8 outflows of Davis
et al. (2008) which we have detected, normalized
to a total of only 5 outflows. The solid curve
is the distribution function expected for a set
of outflows of intrinsic velocity $v_j=100$~km~s$^{-1}$
at random orientations (see equation 1).}
\end{figure}

Particularly, from Figure 11 it is clear that the observed outflows have
a large population in the lowest velocity bin ($0\to 20$ km s$^{-1}$),
for which the ``$v_j=100$ km s$^{-1}$ model'' predicts low frequencies.
This large population of the lowest proper motion velocity bin
therefore implies that some of the outflows in our group indeed
have a low, full spatial velocity,
of at most $v_j \sim 20$ km s$^{-1}$.

\section{Summary and conclusions}

In this paper, we present a first attempt to derive proper motions
of young stellar outflows from Spitzer (IRAC) images. Our analysis
of the NGC 1333 region (which includes the HH 7-11 outflow) shows
that it is indeed possible to do this, which opens up the possibility
of obtaining new ``second epoch'' observations with Spitzer, in order
to derive proper motions for a relatively larger number of IR
outflows.

We have analyzed two epochs at a time interval of $\approx 7$ yr,
and the fact that proper motions are indeed detected can be seen
by taking the difference between the two frames (see Figure 1).
Using these two frames, we first derive the proper motions
of HH 7, 8 and 10.

The resulting proper motions
correspond to plane of the sky velocities of
$\approx (10\pm 4)$ km s$^{-1}$ (see Figure 3 and section 3),
at a distance of 220~pc, which are consistent with the optical
proper motions of these objects derived by Herbig \& Jones (1983)
and Noriega-Crespo \& Garnavich (2001). Our results do not agree
with the $\sim 400$~km~s$^{-1}$ IR proper motions of
Chrysostomou et al. (2000).

We have then derived proper motion velocities for the emission knots
in a $450\times 550''$ field, many of which are aligned with the outflows
described by Davis et al. (2008). For one of these outflows (outflow 3,
see Figure 4), we find a clear bipolar motion (with velocities of
$\sim 80$ km s$^{-1}$). The CO emission of this
outflow is clearly seen in the observations of Curtis et al. (2010).
Another remarkable result is the $\sim 100$ km s$^{-1}$ velocity
derived for the Southern bow shock of outflow 3 (see Figures 4 and 9).

From the proper motions of the 8 outflows of Davis et al. (2008) for which
we have measurements, we obtain a proper motion velocity distribution
function (see section 4.8).
We find that this distribution function is not consistent with
a set of outflows with identical outflow velocities (of
$\sim 100$ km s$^{-1}$) at random orientations. The empirically
derived distribution function indicates that some of the observed
outflows have intrinsic outflow velocities $v_j
\sim 100$ km s$^{-1}$, and that others have a much lower, $v_j\sim
10$-20 km s$^{-1}$ velocity. This consists of a first suggestion
that the velocities $v_j$ (of a group of outflows from recently formed
stars in the same cloud) have a bimodal distribution.

Even though the derived proper motions only give partial constraints
on the kinematics of the outflows in NGC 1333, they do provide some
of the information necessary for modelling this system. This is
of particular interest, since it appears to be a case in which
the outflows have a strong influence on the structure of their
parental cloud, and might even be triggering the formation of
more stars (see Sandell \& Knee 2001). Some other clouds might also
be in a regime of strong perturbation from the outflows from YSOs
(see, e.g., Are et al. 2010; Duarte-Cabral et al. 2012).

It would be interesting to use the kinematics derived from our proper motion
determinations to compute simulations of jet-driven turbulence such
as the ones of Nakamura \& Li (2007) and
Carroll et al. (2010). Would we then obtain a perturbed
cloud such as NGC 1333 (Sandell \& Knee 2001; Arce et al. 2010)?
This would be an interesting complement to the quite extensive
work that has been done on this object (see the papers
quoted above and Quillen et al 2005; Maret et al. 2009; Padoan et al. 2009;
Arce et al. 2010).

\acknowledgements
This research is based in part on observations made with the {\it
Spitzer Space
Telescope} (NASA contract 1407) and has made use of the NASA/IPAC Infrared
Science Archive, both are operated  by the Jet Propulsion Laboratory,
California Institute of Technology, under contract with the National
Aeronautics and Space Administration (NASA).
AR acknowledges support
from the CONACyT grants 60526, 61547, 101356 and 101975.
We thank Luisa Rebull \& John Stauffer for useful conversations about
their program. We also thank an anonymous referee for many corrections
of the first version of this paper.

\appendix{Appendix A: Proper motion mapping}

In order to derive proper motions from pairs of CCD images
it has now become standard to define boxes including emitting
``knots'', and to carry out cross-correlations (between the
2 frames) of the emission within the boxes. The proper motion
is then obtained from a fit to the peak of the cross-correlation
function.
This method has proven to be better than carrying out
direct fits (of e.g., a Gaussian or a 2D paraboloid) to the observed
emission features, because the cross-correlation functions (which
are integrals of the emission within the chosen boxes) have higher
signal-to-noise ratios than the images.

For images with many knots, it is simpler to use a regular array
of boxes (or ``tiles'', as called by Szyszka et al. 2011) for computing
cross correlations. Proper motions for the emission within
each of the boxes are then obtained from fits to their respective
cross-correlation functions.

In this paper, we use tha setup described by Raga et al. (2012).
We define square boxes of side $L$, with spacings of $L/2$ along
each of the coordinates of the images. Therefore, neighbouring
boxes have a superposition region of size $L/2$ along each axis,
as shown in the schematic diagram of Figure 12.

For a given box of size $L$(see Figure 12), we first check whether
or not the condition $f\geq f_{min}$ (where $f$ is the intensity
of the image and $f_{min}$ is a user specified ``minimum flux'') is
satisfied in at least one pixel within a central ``inner box'' of size
$L/2$. If this condition is met at least for a single
pixel in each of the two epochs that are being analyzed, the cross-correlation
function (within the $L$-size box in the two frames) and the proper motion
(from a paraboloidal fit to $3\times 3$ pixels centered on the peak of
the cross correlation function) are computed.

\begin{figure}
\centerline{
\includegraphics[width=200pt,height=200pt,angle=0]{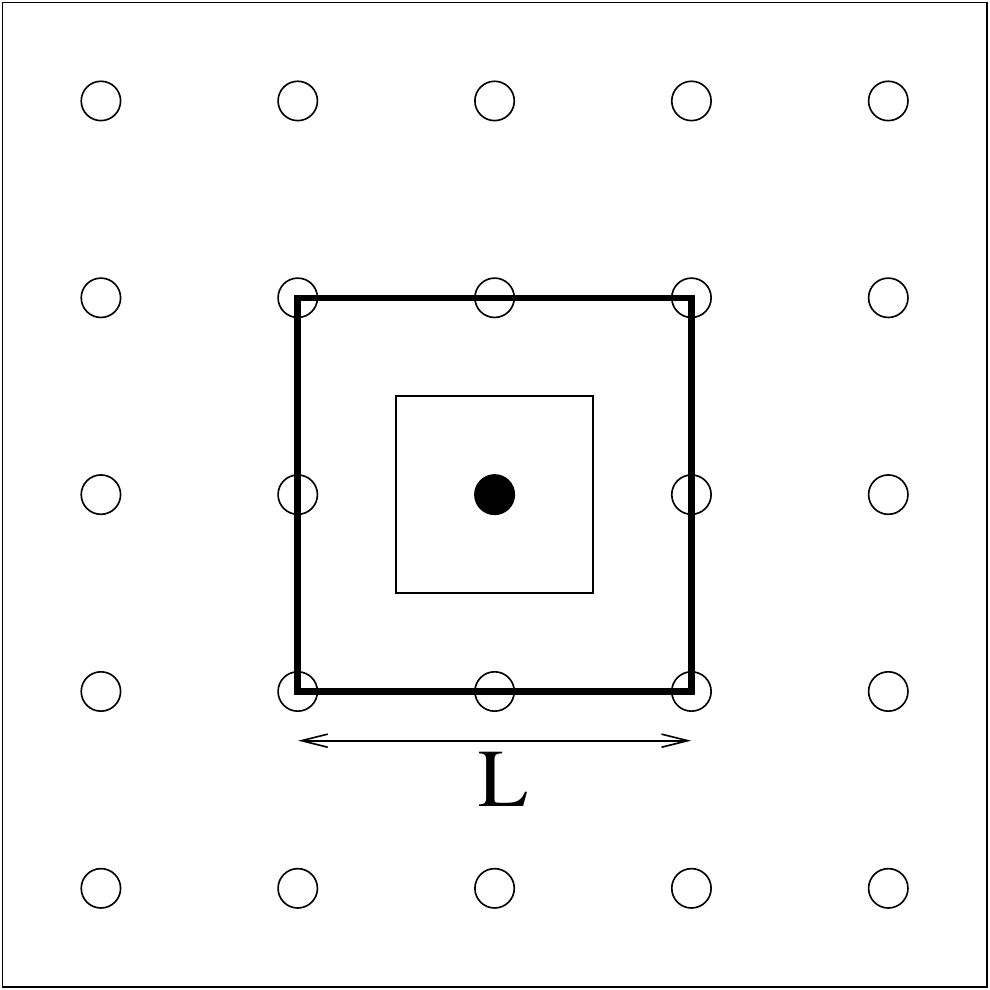}}
\caption{\label{fig12} Schematic diagram showing one of the cross
correlation boxes (thick line box, of size $L$). The
solid circle is the central position of this box, and
the open circles are the central positions of the
nearby cross correlation boxes. The inner box (drawn
with a thin line) of size $L/2$ is used for evaluating
whether or not the peak flux satisfies the criterion
for calculating a proper motion}
\end{figure}

Therefore, in order to compute a ``cross correlation map'', it is then only
necessary to specify the box size ``L'' and the minimum flux $f_{min}$.
Actually, there is another ``hidden'' parameter which is the centering
of the cuadricule of boxes with respect to the pixels of the image. This
is fixed arbitrarily so that the first pixel of the frames that are
being studied coincides with the edge of one of the cross-correlation
boxes.

We find that the peaks of the cross correlation boxes can be fitted to
within $\sim 0.1$~pixel, a precision similar to the one of the alignment
between the two epochs that we have used. Therefore, the formal errors
of our proper motion determinations are of $\sim 0.1$~pixel, corresponding
to a velocity of $\sim 4$~km~s$^{-1}$.

\begin{deluxetable}{rccccc}
\tablecaption{NGC~1333 Observations\label{tbl:colors}}
\tablewidth{0pt}
\tablecolumns{4}
\tablehead{
\colhead{Program ID} & \colhead{Request Key}  & \colhead{Observation Time}
& \colhead{PI} }
\startdata
  6   &  3652864 & 2004-02-10 & Fazio, G.\\
 178  &  5793280 & 2004-09-08 & Evans, N.\\  
61026\tablenotemark{a} & 29323008 & 2011-10-12 & Stauffer, J.\\
      & 29319424 & 2011-11-13 &            \\
\enddata
{\baselineskip=0pt
\tablenotetext{a}{There are 73 independent observations (Request Key)
in this program; the first and last are shown.}
}
\end{deluxetable}

\end{document}